\documentclass[final,5p,times,twocolumn]{elsarticle}

\usepackage{bm}
\usepackage{amssymb}
\usepackage{amsmath}
\usepackage[ruled, vlined, linesnumbered]{algorithm2e}
\usepackage{graphicx}
\usepackage{caption}
\usepackage{subcaption}
\usepackage{indentfirst}
\usepackage{url}
\usepackage{mathptmx}
\usepackage{enumerate}
\usepackage{xcolor}

\journal{Computer Networks}

\begin{document}

\begin{frontmatter}



\title{A Cost-Effective Workload Allocation Strategy for Cloud-Native Edge Services}


\author[inst1]{Valentino~Armani}

\author[inst1,inst2]{Francescomaria~Faticanti}

\author[inst1]{Silvio~Cretti}

\author[inst3]{Seungwoo~Kum}

\author[inst1]{Domenico~Siracusa}

\affiliation[inst1]{organization={Fondazione Bruno Kessler},
            addressline={Via Sommarive, 18, Povo}, 
            city={Trento},
            postcode={38123}, 
            state={Italy}}

\affiliation[inst2]{organization={University of Trento},
            addressline={Trento}, 
            state={Italy}}

\affiliation[inst3]{organization={Korea Electronics Technology Institute},
            addressline={Seongnam-si}, 
            state={Republic of Korea}}

\begin{abstract}
Nowadays IoT applications consist of a collection of loosely coupled modules, namely microservices, that can be managed and placed in a heterogeneous environment consisting of private and public resources. It follows that distributing the application logic introduces new challenges in guaranteeing  performance and reducing costs. However, most existing solutions are focused on reducing pay-per-use costs without considering a microservice-based architecture. We propose a cost-effective workload allocation for microservice-based applications. We model the problem as an integer programming problem and we formulate an efficient and near-optimal heuristic solution given the NP-hardness of the original problem. Numerical results demonstrate the good performance of the proposed heuristic in terms of cost reduction and performance with respect to optimal and state-of-the-art solutions. Moreover, an evaluation conducted in a Kubernetes cluster running in an OpenStack ecosystem confirms the feasibility and the validity of the proposed solution.
\end{abstract}


\end{frontmatter}


\section{Introduction}

The huge amount of data generated by IoT devices and the increasingly stricter latency requirements of applications makes it difficult to rely solely on cloud infrastructures to perform data processing. Indeed, cloud resources usually are physically distant from the source of data and this could lead to high communication delay and bandwidth requirements. To mitigate these drawbacks, Fog Computing enables computation at the edge of the network~\cite{BonomiEtAl_FogAndIoT,YiEtAl_FogComSurvey}.
Fog Computing is a paradigm that brings cloud service features closer to where data is produced~\cite{TorderaEtAL_whatIsfog}. Furthermore, given the limited resources at the edge, a typical fog infrastructure can use private and public resources involving different cloud providers; in such a case, we can refer to it as a \textit{Hybrid Cloud}. A Hybrid Cloud, as defined by the National Institute of Standards and Technology (NIST), is a cloud infrastructure consisting of two or more distinct cloud infrastructures, e.g., private and public~\cite{BadgerEtAl_CloudRecommend}. 

In such an environment the natural trend in the design and development of applications is the microservice paradigm~\cite{NadareishviliEtAl_MicroArchitecture, GanEtAl_ImplicatioofMicroserv}. Microservice applications consist of a collection of loosely coupled modules, namely \textit{microservices}. Each implements specific functionalities and interfaces that, by their composition, provide the intended application functionality. 

One of the main challenges in this context is the minimization of the costs for renting resources in the public cloud while guaranteeing applications requirements \cite{WeintraubCohen_costOptimization}. Indeed, given the shortage of resources at the edge and the microservice-based architecture for applications, a private infrastructure owner is likely to rely on external public clouds to guarantee applications' requirements. The need of \textit{(i)} leveraging hybrid cloud models reducing pay-per-use costs, and \textit{(ii)} spatially distributing the computational capacity to satisfy applications' constraints requires an advancement of current solutions usually adopted in Cloud Computing. In particular, in the case of a distributed and heterogeneous computing infrastructures composed of different \textit{regions}, e.g. one or more public clouds of different vendors, one or more private clouds and edge regions, it is crucial to decide where to deploy a given microservice and when it is better to re-configure the status of such a deployment. Several benefits can be obtained by these decisions: \textit{(i)} optimization of the usage of private resources, \textit{(ii)} optimization of the cost of public resources, and \textit{(iii)} optimization of the applications' performance. Several allocation strategies are presented in the cloud computing literature~\cite{wolke2014planning}, however the hybrid cloud scenario requires to take into account additional elements with respect to the traditional cloud: the bandwidth and the latency between different regions and costs of applications deployment on public clouds.

This paper studies a workload orchestration strategy able to minimize the costs of a pay-per-use public cloud while taking into account application requirements and performance. Considering a CPU-intensive application, we aim to demonstrate how it is possible to meet these objectives by redirecting the data flow (i.e., the requests received by the application) from one region to another.

\textit{Main contribution}. In this paper we investigate \textit{how a cloud-cost-aware strategy  impacts the performance of a microservice-based application in a hybrid cloud environment.} In order to explore such a scenario we formulated the problem of cost optimization that results equivalent to an instance of the well known Virtual Network Embedding (VNE) problem~\cite{CaoEtAl_VNEsurvey}. Given the high computational complexity of the problem, known to be NP-hard \cite{FischerEtAl_VNEsurvey}, we proposed a novel efficient heuristic solution. This new solution is driven by two main criteria: the cost reduction and the load balancing between microservices to satisfy the application's requirements. The heuristic solution is based on the concept of \textit{replicasArray} that will efficiently distribute the traffic within the cloud clusters. This concept will be introduced in Section~\ref{sec:algo}. In order to validate the proposed solution, we conducted a simulation study of the heuristic and its parameters comparing the proposed solution against the optimal solution and a state-of-the-art solution. Finally, a real implementation of such an approach is presented.

The remainder of this paper is structured as follows. The next section presents related works. Section~\ref{sec:model} and Section~\ref{sec:form} introduce the system model and the problem formulation, respectively. The algorithmic solution is described in Section~\ref{sec:algo} and its numerical evaluation is presented in Section~\ref{sec:eval}. Section~\ref{sec:impl} describes the performance of the proposed solution in a real scenario. A concluding section ends the paper.

\section{Related work}

This work addresses the problem of workload allocation and load balancing in a hybrid/fog environment. Several works have been presented in the literature of the broad research area of cloud computing.

In \cite{WangAdaptiveSF} the authors proposed the Adaptive-Scheduling-with-QoS-Satisfaction algorithm that aims to maximize the utilization of the private cloud and minimize the cost of the public cloud when users submit their computing jobs in a hybrid cloud environment. To achieve that, they exploited run-time estimation and several fast scheduling strategies for near-optimal resource allocation. However, this work does not fit the microservice paradigm since it is focused on the specific concepts of `jobs' and `tasks'. A `task' is the basic unit of a user request that can be handled and processed, one at a time, by a computation resource. A `job' is a set of tasks. 

Another study that utilizes the notions of `jobs' and `tasks' is the one presented by Shah-Mansouri and Wong in \cite{MansouriWong_offloading_game}. In this paper, they studied the allocation of fog computing resources in a hierarchical environment comprehensive of fog and remote cloud computing services. They formulated a computation offloading game to model the competition between users, which aims to maximize their quality of experience (QoE), and allocate the limited resources available. Due to the high computational complexity required by their solution, they proposed a near-optimal resource allocation algorithm that executes in polynomial time. In this work, however, they investigated neither public cloud environment nor monetary costs.

A study of the economic aspects of a joint vertical and horizontal cloud federation is presented in \cite{ChenEtAl_CloudFederation}. In this paper, Chen et al. analyzed the interrelated workload factoring and federation formation among secondary clouds. Moreover, they provided a scalable algorithm that can help cloud providers to optimally select partners and outsource workload. This work, differently from our approach, assumes a predictable requests pattern and achieves costs reduction by sharing resources among cloud providers.

In \cite{DengEtAl_WorkAllocation_power} the authors formulated a workload allocation problem that aims to find an optimal workload allocation that minimizes power consumption while having constraints on the service delay. They modeled the power consumption function and the delay function and formulated the workload allocation problem in a fog computing infrastructure. Although related, this paper addresses the problem of cost reduction from a different perspective with respect to ours; the costs we consider are not due to power consumption of all the available private servers but to a pay-to-use public cloud.

Maswood et al. in \cite{MaswoodEA_BandwidthLoadbalancing} proposed an optimization model in a three-layer fog-cloud computing environment. This model has a composite objective function that aims to minimize the bandwidth cost and to provide load balancing, both in terms of links’ bandwidth and of servers’ CPU. Differently from our approach, this work considers usage-based costs related to the bandwidth consumption without taking into account the microservice paradigm nor any Quality-of-Service requirements.

Concerning load balancing problems, Yu et al. in \cite{YuEtAl_loadb_micro} studied the problem of load balancing across microservice instances taking into account inter-dependencies among modules. They firstly presented a graph-based model for describing the load dependencies and then formulated a QoS-aware load balancing problem. Given its NP-hardness, they proposed a fully polynomial-time approximation scheme. However, the formulation presented here does not allow a microservice instance to communicate with more than one instance of a different microservice. Thus, differently from our approach, the proposed model does not allow a service to send traffic to all the available replicas of a different module.

The workload allocation problem is firstly tackled from the perspective of a Virtual Network Embedding (VNE) problem. VNE is used to map virtualized networks on a substrate infrastructure, optimizing the layout for service-relevant metrics \cite{FischerEtAl_VNEsurvey}. VNE deals with the allocation of virtual resources addressing two sub-problems: allocating virtual nodes in physical nodes and mapping virtual links connecting virtual nodes to paths connecting the corresponding physical nodes in the substrate network. \cite{FischerEtAl_VNEsurvey}. The output of such a problem is an optimal mapping of microservices and traffic on the cloud infrastructure that satisfies applications and cloud provider requirements. However, Virtual Network Embedding problem is known to be ${NP}$-hard \cite{ChowdhuryEtAl_CoordinatedVNE} \cite{YuEtAl_RethinkingVNE} and most of the resolution methods are based on heuristic solutions \cite{CaoEtAl_VNEHeuristicSurvey}. 

In this work, the workload allocation strategy is implemented in a reactive fashion. Reactive approaches represent one of two possible techniques, together with the proactive one, used to implement \textit{auto-scaling} \cite{Yazdanov_TowardsAutoScaling} \cite{ParminderEtAl_AutoScaling}. Auto-scaling is a method used in cloud computing to dynamically add or remove application resources according to workload demand and performance constraints. Reactive solutions respond only on the current state of the application and, usually, check whether a set of metrics exceeds a given threshold and adapt resources depending on demands. On the other hand, in order to provide appropriate resources beforehand, proactive solutions perform scaling actions based on future resource demands forecast. However, due to their simplicity and adequate performance, many companies and organizations still use reactive approaches \cite{LiuEtAl_FuzzyBasedAutoScaler} \cite{LoridoEtAl_ReviewAutoScaling}. 

\section{System Model}
\label{sec:model}
\subsection{Assumptions}
\label{sec:sysDesign}
We outline the system design of a hybrid cloud where a cloud service provider aims to achieve cost-reduction while trying to meet users Quality-of-Service (QoS) expectations. The goal is to provide a workload orchestration strategy that minimizes the costs of the pay-to-use public cloud while taking into account hardware capabilities and application requirements. This is done by scheduling the application and routing its internal workload in a way to minimize the provider costs while guaranteeing processing and communication delays below given thresholds.

The infrastructure we are considering is a hybrid cloud composed of two distinct cloud infrastructures; a public and a private cloud. Moreover, in the hybrid cloud, we can identify several and heterogeneous \textit{regions} or, equivalently, \textit{fog regions}.  A region is a specific geographical location where resources are placed. We assume that regions can have different hardware and that the computation capabilities at the edge are reduced with respect to the ones available in a central or in a public region.

The hybrid cloud runs a container orchestration system, e.g. Kubernetes, and hosts an application that offers to users some functionalities. The application is developed following the microservice paradigm and its modules are placed within the system. To reduce latency, leverage data locality and limit the internal bandwidth usage, the cloud service provider aims to take full advantage of edge regions bringing processing closer to the source of data. This is done by deploying in the fog regions a subset of the microservices that compose the application. To further decrease the latency, replicas of some of the services are placed over the different regions available on the infrastructure. The network functionalities and microservices communication are managed by a dedicated infrastructure layer made of a collection of network proxies, i.e. a service mesh.

We assume that under low/normal workload conditions, when the number of requests to process is not too high, the computation can be handled by the edge resources only. This allows the provider to keep latency low while ensuring privacy. However, as the workload increases, edge resources may not be enough to meet users Quality-of-Service and a fog region may be saturated. For this reason, application completion time and cluster resources are constantly monitored. When a time constraint violation is going to happen the service mesh, leveraging the monitoring set-up, is instructed to redirect part of the traffic to a replica hosted in a different private region of the cloud; the load quantity each region receives depends on its residual capacity. Then, if the workload saturates all the private resources, requests are also routed to the pay-to-use public cloud. Finally, when the load decreases, the service mesh is instructed to stop sending traffic to the public cloud and, if conditions allow it, handle all the computation by edge resources only.

This workload orchestration allows the cloud service provider to host a responsive application that leverages private resources while using the public cloud only on heavy workload conditions. Moreover, the load redirection based on residual capacity can handle different requests patterns and manage heterogeneous hardware; traffic quotas for each region are computed dynamically at run-time and based on the resources available. The cost-reduction is ensured by the fact that the public cloud is pay-to-use and we suppose it has no costs if it is not receiving traffic, which is the case for most serverless platforms \cite{ServerlessSurvey}. We do not examine user mobility and we assume that requests submitted to the application, always coming from the same location, require different but constant access delay to reach each region. Finally, we look at the regions as a whole and we do not consider single servers within them.

\subsection{System model}
\begin{table}[t]
    \begin{tabular}{ c p{6cm}} 
     \hline
     \textit{Symbol} & \textit{Meaning}  \\ \hline
     $\mathcal{R}$ & Set of regions $|\mathcal{R}|=R$  \\ 
     $\bm{C_r}$ & Memory, storage and processing capacity of region $r$: $\bm{C_r}=(C_r^M,C_r^P,C_r^S)$ \\ 
     $\mathcal{M}$ & Multiset of microservices instances $\mathcal{M}= \cup_{i=1}^R M_i$ \\
     $f: \mathcal{M} \rightarrow \mathbb{R}^+$ & Public cloud cost function \\
     $G=(V,E)$ & Network architecture graph \\
     $ \Gamma =(N,L)$ & Application architecture graph \\
     $\bm{c_m}$ & Memory, storage and processing requirements of microservice $m$: $\bm{c_m}=(c_m^M,c_m^P,c_m^S)$ \\
     $\omega_m$ & CPU instructions required by microservice $m$ to process one request\\ 
     $\mathcal{P}$ & Set of paths $p$ between any couple of regions \\
     $P_{i,j} \subseteq \mathcal{P}$ & Set of paths between regions $i \in \mathcal{R}$ and $j \in \mathcal{R}$ \\
     $ x_{m,r} \in \{0,1\}$ & Boolean variable indicating microservice $m$ deployed in region $r$ \\
     $ y_p^{(u,v)} \in \{0,1\}$ & Boolean variable indicating link $(u,v) \in L$ mapped to path $p$  \\
     $ \pi: \mathbb{R} \rightarrow \mathbb{R}^+$ & Processing rate function, measured in MIPS\\
     $C_r^{'P}$ & Residual processing capacity of region $r$\\
     $\Delta_m$ & Number of requests that a microservice $m$ has to process\\
     \hline
    \end{tabular}
    \caption{Notation used in the formulation.}
\end{table}

A hybrid cloud infrastructure consists of a private fog and a public cloud infrastructure which are deployed over a set of geographic regions $\mathcal{R}=\{1, ..., R\}$. The private regions are organized in a layered architecture while the public one runs off-premise and we denote it as $r_{pub}$. Each region $r \in \mathcal{R}$ is characterized by a capacity vector $\bm{C_r}=(C_r^M,C_r^P,C_r^S)$ and a residual vector $\bm{C_r^{'}}=(C_r^M,C_r^P,C_r^S)$ representing the memory, the processing and the storage resources available and residual, respectively, in $r$. The processing rate function $\pi$ of a region $r$ is defined linearly w.r.t. the residual capacity  $\bm{C_r^{'P}}$; the less the residual capacity is, the lower the function value is. We assume that individual resources available at the edge of the fog computing infrastructure are significantly smaller than the ones in the core region. A region $r$ can host a set of microservices, $M_r$, of an application. We say that microservice $m$ “belongs” to a given region because an instance of that microservice is placed there. We assume that the public region has a linear cost depending on the number of microservices active in it and we define $f$ as the public cloud cost function. 

\textit{Network Model.} Private regions are interconnected and can communicate with the public cloud. Network communication is subject to delays and bandwidth constraints. In this work we consider static network delays and bandwidth as commonly considered in the literature~\cite{Vaezpour_PerformanceAS}. We model the hybrid cloud system as a directed weighted graph $G=(V,E)$ where $V=\mathcal{R}$ and $E \subseteq V \times V$. Each link $(i, j) \in E$ has a weight $(d_{i,j},b_{i,j})$, i.e., the delay and the bandwidth, respectively, of $(i, j)$. The set of paths $p$ between any couple of regions is denoted as $\mathcal{P}$ while $P_{i,j} \subseteq \mathcal{P}$ represents the set of path between regions $i \in \mathcal{R}$ and $j \in \mathcal{R}$.

\textit{Application Model.} An application is a set of microservices. It is modeled as a directed weighted graph $\Gamma =(N,L)$ where $N$ is the set of nodes, denoting the microservices, and $L$ is the set of links, representing directed API calls between microservices, $L \subseteq N \times N$. The weight of each link $(u,v) \in L$ is given by the maximum delay that this link can tolerate, $d_{u,v}^{max}$, and the maximum throughput generated from $u$ to $v$, $\lambda_{u,v}$. Each microservice $m \in N$ has memory, processing and storage requirements $\bm{c_m}=(c_m^M,c_m^P,c_m^S)$. Moreover, each microservice $m$ is characterized by $\omega_m$, the CPU requirements to process one request in input, and by $\Delta_m$, the average size of all the incoming requests that it had to process.

\section{Problem formulation}\label{sec:form}
The workload allocation problem is addressed from the perspective of cloud service providers. They aim to leverage localized data processing and minimize their overall costs subject to a maximum completion time. Indeed, the public region has costs depending on its utilization. This problem can be seen as an instance of the Virtual Network Embedding (VNE) problem that has been proven to be NP-hard\cite{BrogiEtAl_DeployFogApp}. 

We formulate the problem as an integer programming problem as follow:

\textit{Decision variables.} We define two Boolean variables:
\begin{equation}
    x_{m,r} =
    \begin{cases}
      1 & \parbox[t]{.35\textwidth}{if a microservice $m \in N$ is deployed within region $r \in \mathcal{R} $}\\
      0 & \text{otherwise}
    \end{cases}
    \nonumber
\end{equation}
to assign application microservices to regions, and:
\begin{equation}
    y_p^{(u,v)} =
    \begin{cases}
      1 & \text{if the link $(u,v) \in L$ is mapped to path $p \in \mathcal{P} $}\\
      0 & \text{otherwise}
    \end{cases}
    \nonumber
\end{equation}
$\forall (u,v) \in L, \forall p \in \mathcal{P}$,  to assign links of the application to physical paths. 

\textit{Objective function.} The objective is to minimize the cost from using the public cloud. We denote as $\psi$ and $\zeta$ the processing and the communication delays upper bounds, respectively. Therefore, the integer program writes:
\begingroup
\allowdisplaybreaks
{\small
\begin{alignat}{2}
  & \text{minimize:} \quad \quad   f (\sum_{m \in N} x_{m,r_{pub}}) & \quad & \label{eq:main} \\
  & \text{subject to:}  & & \nonumber \\
  &                     \sum_{m \in N} x_{m,r} \bm{c_m} \leq \bm{C_r}, &  & \forall r \in V \\
  &                     \sum_{(u,v) \in L} \sum_{p \in \mathcal{P}:(i,j) \in p} \lambda_{u,v}y_p^{(u,v)} \leq b_{i,j} &  & \forall (i,j) \in E \\
  &                     \sum_{(i,j) \in p} d_{i,j} y_p^{(u,v)} \leq d_{u,v}^{max} &  & \forall (u,v) \in L, \forall p \in \mathcal{P} \\
  &                     z^{(u,v)}_{i,j} = x_{u,i} \cdot x_{v,j} &  & \forall i \in V, \forall j \in V, \forall (u,v) \in L \\ 
  &                     t^{(u,v)}_{i,j} \leq (1-x_{v,i}) + (1-x_{u,j}) &  & \forall i \in V, \forall j \in V, \forall (u,v) \in L \\ 
  &                     t^{(u,v)}_{i,j} \geq (1-x_{v,i}) &  & \forall i \in V, \forall j \in V, \forall (u,v) \in L \\ 
  &                     t^{(u,v)}_{i,j} \geq (1-x_{u,j}) &  & \forall i \in V, \forall j \in V, \forall (u,v) \in L \\ 
  &                     \sum_{p \in P_{i,j}} y_p^{(u,v)} \leq z^{(u,v)}_{i,j} &  & \forall i \in V, \forall j \in V, \forall (u,v) \in L \\
  &                     \sum_{p \in P_{i,j}} y_p^{(u,v)} \leq t^{(u,v)}_{i,j} &  & \forall i \in V, \forall j \in V, \forall (u,v) \in L \\
  &                     \sum_{p \in P_{i,j}} y_p^{(u,v)} \geq z^{(u,v)}_{i,j} + t^{(u,v)}_{i,j} - 1 &  & \forall i \in V, \forall j \in V, \forall (u,v) \in L \\
  &                     \sum_{r \in V} x_{m,r} \geq 1 &  & \forall m \in N \\
  &                     d_m \leq \psi_m &  & \forall m \in N \\
  &                     \sum_{p \in \mathcal{P}} \sum_{(u,v) \in L} \sum_{(i,j) \in p} d_{i,j} y_p^{(u,v)} \leq \zeta &  & \\
  &                     x_{m,r} \in \{0,1\} &  & \forall (m,r) \in N \times \mathcal{R}\\
  &                     y_p^{(u,v)} \in \{0,1\} &  & \forall (u,v) \in L, \forall p \in \mathcal{P} \\
  &                     z^{(u,v)}_{i,j} \in \{0,1\} &  & \forall (u,v) \in L, \forall (i,j) \in E \\
  &                     t^{(u,v)}_{i,j} \in \{0,1\} &  & \forall (u,v) \in L, \forall (i,j) \in E
\end{alignat}
}
\endgroup
where:
{\small
\[ d_m =  \frac{\Delta_m \cdot \omega_m}{\sum_{r \in V} x_{m,r} \cdot \pi(C_r^{'P})} \]
\[C_r^{'P} = C_r^{P} - \sum_{m \in N} c_m^p x_{m,r}, \forall r \in V\]
\[\sum_{m \in N}  \psi_m =  \psi\]
}
Constraint (2) bounds resource utilization on regions in terms of memory, processing and storage capacity. (3) and (4) are constraints on the edges' bandwidth capacity and on the application' links delay, respectively. Constraints (5)-(11), that can be also written as $\sum_{p \in P_{i,j}} y_p^{(u,v)} = x_{u,i} \land x_{v,j} \land (\neg x_{v,i} \lor \neg x_{u,j})$, guarantee that the link between two microservices, that should communicate and are in two different regions, is mapped to a physical path.
By constraint (12), we ensure that all microservices of the application are deployed. Constraint (13) captures that the processing time of each microservice can not exceed the related threshold, $\psi$. A microservice processing delay, $d_m$, is defined as the computing requirement over the processing rate obtained as a function of the residual computing capacity of each region where $m$ is placed. Finally, constraint (14) ensures that the overhead due to the communication delay is below a given threshold $\zeta$.

\section{Algorithmic solution}\label{sec:algo}
The integer problem defined in~\eqref{eq:main} should be solved each time the number of outstanding requests varies significantly. Moreover, the high computational complexity of the problem, due to the NP-hardness of VNE, does not allow efficiently computing a solution of the problem as the input size increases. 
Hence, we propose a heuristic algorithm whose output is computed in reasonable time. The heuristic algorithm consists of two stages. The first one addresses the problem of defining the required structure and creating an initial placement of the application microservices. Then, in the second stage, the algorithm performs a workload allocation which aims to minimize the provider costs while trying to respect the maximum application completion time.

\subsection{Initial placement}\label{sec:placement}

This stage introduces the concept of \textit{replicasArray}, which is at the core of the workload redirection, and proposes a baseline algorithm to deploy the application microservices within the infrastructure. Hence, this procedure is executed only once, at the beginning. The output is a mapping of the placement of the microservices and their replicas onto the available physical resources. 

Given the set of instances of microservices that must be deployed, the \textit{initial placement} procedure consists of the following steps:
\begin{enumerate}
    \item \textit{Sorting of the application graph}: an order of the microservices that constitute the app is established performing a Breadth-First Search traversal of the graph. The visit starts from the first node that users reach when sending requests to the application; this allows us to capture how workload flows within the application.
    \item \textit{Sorting of the infrastructure regions}: an order is obtained sorting the regions depending on users' access delay. This step allows the algorithm to consider users' propagation delay to reach the cloud and to leverage data processing closer to users. Moreover, to reduce costs, all private regions come before than the public ones.
    \item \textit{Node and link mapping}: the algorithm iterates on the set of instances of microservices ordered depending on step 1. For each microservice the first region, ordered as defined in step 2, that satisfies the following criteria is selected:
    \begin{enumerate}[i)]
        \item It has enough residual resources to host the microservice.
        \item Its residual CPU capacity after the deployment of the microservice is above a fixed threshold, $\tau$.
    \end{enumerate}
    Criteria \textit{i} and \textit{ii} ensure that a region can host a microservice and without becoming saturated. Link mapping is performed in conjunction with node mapping. If two microservices that need to communicate are placed on two different regions, the existence of a path that satisfies the delay and bandwidth constraints is ensured. In the case of multiple feasible paths between regions, the algorithm selects the least congested one.
\end{enumerate}

Once an instance of every microservice is deployed, to make workload allocation possible, the algorithm replicates some of the microservices doing the following:

\begin{enumerate}
 \setcounter{enumi}{3}
    \item \textit{ReplicasArray definition}: a \textit{strict subset} of microservices that constitute the app is selected. Given the order defined by the graph traversal obtained in step 1, this subset is made of consecutive microservices. These microservices constitute what we call a \textit{replicasArray}. A replicasArray will be deployed, i.e. replicated, in each region that satisfies defined constraints. The number of microservices to include is application-dependent and, ideally, should be large enough to meet time constraints even under high workload conditions while as small as possible to minimize resource occupation. To perform the selection, given the application graph traversal and starting from the last leaf according to the order of step 1, search for the \textit{strict subset} of the chosen length that has the \textit{highest CPU demand}. This subset must contain consecutive microservices to let the traffic flow among them without involving microservices in other regions, thus avoid creating an overhead on links between regions due to backward and forward requests. Indeed, having replicas of consecutive microservice allows us to introduce, in the application flow, a single point of traffic-splitting. Moreover, we select the subset with the highest CPU demand to replicate the most computationally expensive microservices.
    
    \item \textit{ReplicasArray placement}: the algorithm iterates on the set of the regions ordered according to step 2. In each region that satisfies the following criteria a replicasArray is placed:
    \begin{enumerate}[I)]
        \item It has enough residual resources to host the replicasArray.
        \item Its residual CPU capacity after the deployment of the replicasArray is above a fixed threshold, $\tau$.
        \item It does not already contain a microservice of the replicasArray set.
        \item There exists a path that satisfies delay and bandwidth constraints of the incoming and outgoing links of the replicasArray's microservices.
    \end{enumerate}
     Criteria \textit{I, II} are defined similarly and for the same motivations as step 3. Criterion \textit{III} avoids deploying more than one instance of the same microservice on the same logical unit. Criterion \textit{IV} ensures that a feasible path that allows the communication to and from the microservices of a replicasArray is present.
\end{enumerate}

\begin{figure}[t]
    \includegraphics[width=0.48\textwidth]{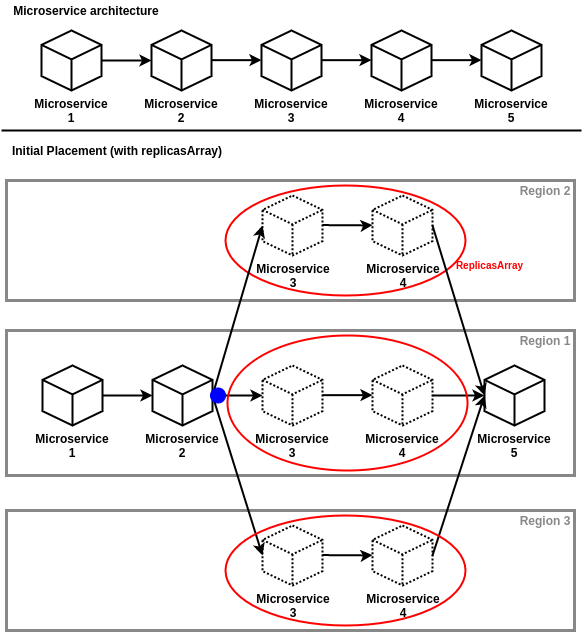}
    \centering
    \caption{Example microservice architecture diagram and \textit{initial placement} of the application setting the replicasArray size to two.}
    \label{fig:replicasArray}
\end{figure}

Figure \ref{fig:replicasArray} depicts the \textit{initial placement} of an example application in a hybrid cloud made of three regions and assuming a replicasArray dimension of two. A replicasArray instance, represented as the subset of dashed microservices circled in red, is deployed in each region. The blue dot corresponds to the (single) point of traffic splitting. 

\subsection{Data Stream Reconfiguration (DSR)}

\begin{algorithm}[t]
    \DontPrintSemicolon
    \SetAlgoLined
    \SetKwInOut{Input}{Input} \SetKwInOut{Output}{Output}
    
    \Input{$G=(V,E)$, $ \Gamma =(N,L)$, \textit{map}}
    \Output{Reconfiguration of the workload}
    \BlankLine
    
    \textit{activeRA} $\leftarrow$ \{\} \tcp*[f]{activeReplicasArray} \;
     \textit{clusterSort} $\leftarrow$ sortIncreasingDelay($G$, $r$) \;
    
    \While{TRUE}
    { 
       \textit{decision} $\leftarrow$ isWorkloadShiftRequired( ... ) \;
        \If{(\textit{decision} == ACTIVATE) OR (\textit{decision}~$==$~DEACTIVATE)}
        {
            \eIf{\textit{decision} == ACTIVATE}
            {
                \textit{activeRA} $\leftarrow$ activeNewReplicasArray(\textit{clusterSort, map, activeRA}) \;
            }
            {
                \textit{activeRA} $\leftarrow$ removeLastElement(\textit{activeRA})
            }
            \textit{totResidualCPU} $\leftarrow$ 0 \;
            \ForEach{region in activeRA} 
            {
                \textit{totResidualCPU} $\leftarrow$ \textit{totResidualCPU + } residualCPUCapacity\textit{(region)} \;
            }
            \ForEach{region in activeRA} 
            {
                \textit{trafficQuantity} $\leftarrow$ residualCPUCapacity\textit{(region)  / totResidualCPU} \;
            }
             \tcc{Reconfigure stream according to \textit{trafficQuantity}}
        }
    }
    \caption{Data Stream Reconfiguration (\textit{DSR})}
    \label{alg:workloadShift}
\end{algorithm}

\begin{algorithm}[t]
    \DontPrintSemicolon
    \SetAlgoLined
    \SetKwInOut{Input}{Input} \SetKwInOut{Output}{Output}
    
    \Input{\textit{lastDecision, lastRequests, currentRequests, currentProcesTime}}
    \Output{A decision about the redirection of the traffic}
    \BlankLine
    
    \textit{decision} $\leftarrow$ \textit{NONE}
    
    \If{\textit{currentProcesTime} $\geq$ upperThreshold}
    {
       \textit{decision} $\leftarrow$  \textit{ACTIVATE}
    }
    \If{\textit{currentProcesTime} $\leq$ lowerThreshold}
    {
       \textit{decision} $\leftarrow$  \textit{DEACTIVATE}
    }
    \BlankLine
    \If{\textit{(lastDecision} $==$  \textit{NONE) OR  (lastDecision}~$==$~DEACTIVATE)}
    {
       \KwRet{decision}
    }
    \ElseIf{\textit{lastDecision} $==$ ACTIVATE}
    {
        \If{(decision $==$ ACTIVATE) OR (decision~$==$~NONE)}
        {
            \KwRet{decision}
        }
        \ElseIf{decision $==$ DEACTIVATE}
        {
            \eIf{currentRequests $>$ (lastRequests - $q\%$)}
            {
                 \KwRet{NONE}
            }
            {
                 \KwRet{decision}
            }
        }
    }
    \caption{isWorkloadShiftRequired procedure}
    \label{alg:isWorkloadShiftRequired}
\end{algorithm}

\label{sec:heu_workload_redirection}
After the placement, the workload needs to be managed in order to avoid exceeding the maximum allowed completion time while minimizing the provider costs. This is done by redirecting the load to different replicasArrays depending on the residual capacity of the host region. In detail, Algorithm \ref{alg:workloadShift} performs, periodically, the following steps: 
\begin{enumerate}
    \item \textit{Is workload shift required?}: we check this at line 4. A decision about traffic redirection to satisfy the processing time constraint is taken. This depends on the processing time delay, which is related to current users' requests and regions' capacity, and the last redirection allowed. The procedure, described in Algorithm \ref{alg:isWorkloadShiftRequired}, defines an upper and a lower threshold on the processing delay; whatever these values are exceeded a ``memory" based method is triggered to reach a decision. This technique leverage the knowledge of the last redirection to avoid taking a succession of opposite decisions while having similar workload conditions. In fact, using only a threshold-based methodology may cause oscillation problems \cite{LiuEtAl_FuzzyBasedAutoScaler}. In detail, once a threshold is exceeded, Algorithm \ref{alg:isWorkloadShiftRequired} formulates a decision (lines 2-7), a candidate one, and checks at line 8 the previous allowed one. If the algorithm has not already taken a decision or the last allowed one is ``deactivate a replicasArray" the candidate decision is returned at line 9. Otherwise, if the last allowed decision is ``activate a new replicasArray", line 11, we apply the ``memory" based technique. If the candidate decision is the same or $NONE$ the decision is returned (lines 12-13). If it is not the case and the number of requests is not decreased above a percent quantity \textit{q}, the candidate decision is rejected in favour of a "shift not necessary" decision. This is visible from lines 15 to 17. Otherwise, the procedure returns the candidate decision, line 19.
    
    If Algorithm \ref{alg:isWorkloadShiftRequired} returns decision ``redirection of the traffic not required", Algorithm \ref{alg:workloadShift} skips the following steps, 2 and 3. The ``memory" based method is specialized only for the ``activation" case since having it also on the ``deactivate" one does not allow the algorithm to be reactive to decisions that cause the exceeding of the upper threshold. In this way, instead, the algorithm is responsive to time constraint violations while removing instability. 
    
    \item \textit{Activation/deactivation of a replicasArray}: based on the decision of the previous procedure, Algorithm \ref{alg:workloadShift} tries to modify the active replicasArrays accordingly. This step activates replicasArrays following the order of the regions defined running the procedure \textit{Sorting of the infrastructure regions}, line 7, and deactivates them following the reverse order, line 9. The activation or deactivation involves one replicasArray at a time. If no replicasArray is available, to be activated or deactivated, the decision taken in the previous step does not involve any modifications. Following the sorting of the regions based on users’ access delay, where public regions come after the private ones, allows to leverage data processing closer to users while reducing costs.
    
    \item \textit{Traffic splitting:} the definition of replicasArray allows to activate several replicas while having a single point of traffic splitting. Indeed, the outgoing traffic of the service placed before the replicasArrays is managed to be distributed among the available replicas. Given the residual CPU capacity of all the regions hosting the activated replicasArrays (line 13), the quantity of requests that each replicasArray, $i$, receives is given by:
    {\small
    \begin{equation}
        quantity_i = \frac{res\_cpu\_capacity\_hostregion(i)}{\sum\limits_{ar \in activeReplicasSet} res\_cpu\_capacity\_hostregion(ar)}
    \end{equation}
    }
    where $i \in activeReplicasSet$ and $quantity_i$ is the percentage of $\Delta_m$ of the first microservice of the replicasArray. This is visible at lines 15-16. This formulation provides load balancing since the quantity of workload that each replicasArray receives depends on the residual capacity of the region that hosts it.  
\end{enumerate}

\textit{Computational complexity}. We look at the computational complexity of the \textit{DSR} algorithm. The cluster sort procedure, executed only once before the main loop, has $\mathcal{O}(|V|\log|V|)$ time complexity assuming an efficient sorting algorithm. The procedures isWorkloadShiftRequired, activeNewReplicasArray, removeLastElement and residualCPUCapacity have constant time complexity. The cardinality of \textit{activeRA} is $\mathcal{O}(|V|)$ in the worst case. Consequently, the body of the endless loop takes $\mathcal{O}(|V|)$ time. This confirms the polynomial complexity of the procedure.

\section{Evaluation} \label{sec:eval}
\label{cha:evaluation}
In this section we evaluate the presented heuristic solution with respect to the defined goal: minimize the public cloud costs subject to time constraints. Firstly, we compare the heuristic algorithm against the optimal solution provided by the mathematical model. The MILP formulation provides a lower bound to the provider’s costs; this allows us to evaluate the heuristic solution and to inspect how its parameters influence the results. Then, to understand how our proposal is good in the defined goal and to test its performances, leveraging  \cite{MaswoodEA_BandwidthLoadbalancing}, we compare our solutions to a load-balancing workload allocation strategy.

In \cite{MaswoodEA_BandwidthLoadbalancing} Maswood et al. (MEA) proposed an optimization model in a cooperative three-layer fog-cloud computing environment. They considered task offloading from fog to a cloud or a public cloud on demand. The model, formulated as Mixed-Integer Linear Programming (MILP) problem, has a composite objective function to minimize the bandwidth cost and provide load balancing. They assigned weight factors to each component of the objective function; by modifying their values it is possible to recreate several workload allocation strategies. The considered goals and their related weights are: to minimize bandwidth cost, $\alpha$, to minimize maximum link utilization of network links, $\beta$, and to minimize the maximum server resource utilization, $\gamma$. By setting the bandwidth parameters to values that associate a higher cost to reach the public region, their minimize bandwidth cost goal can be matched to our minimize on-demand costs goal.

\subsection{Simulation settings}
\label{sec:eval_simulation_settings}
The heuristic algorithm is developed as a Python program, while the integer programming model~\eqref{eq:main}, and the literature solution, shown in \cite{MaswoodEA_BandwidthLoadbalancing}, are implemented with the Gurobi Python API version 9.0 \cite{gurobi}. The parameters for the topology and the application are set assuming possible values of a modern fog infrastructure and microservices-based application \cite{PallewattaEtAl_PlacementFog}.

\textit{1) Topology:} the hybrid cloud is modelled as a fully connected graph with two private regions, an edge and a core one,  and a public region. The considered cloud computing resources have values within predefined intervals and depending on their type. More details in Table \ref{tab:topologyReources}.
{\small
\begin{table}[t]
\footnotesize
\begin{tabular}{|c|c|c|c| } 
 \hline
 Resource & Edge Region & Central Region & Public Region\\ \hline
 Memory (GB) & [4-16] & [6-16] & [8-32] \\ 
 CPU (MIPS) & [2000-4000] & [4000-6000] & [4000-8000] \\
 Storage (GB) & [20-100] & [20-100] & [40-100] \\
 Bandwidth (Mbps) & [400-800] & [400-800] & [400-1000] \\
 Delay (ms) & [10-50] & [20-50] & [30-100] \\
 \hline
\end{tabular}
\caption{Cloud computing resources capabilities.}
\label{tab:topologyReources}
\end{table}
}
We set a maximum amount of resources for the public clouds due to monetary constraints for renting them and we ensure fewer computation capabilities at the edge region. Moreover, we assume that links between private regions introduce lower delays than links involving public resources.

\textit{2) Application:} the application is modelled as a directed graph consisting of five microservices. The values characterizing each component are within predefined ranges, as specified in Table \ref{tab:MicroservicesRequirements}. The maximum delay tolerated was set to the constant value of 100 ms. 

\begin{table}[t]
\begin{center}
\begin{tabular}{ |c|c| } 
 \hline
 Requirement & Values \\ \hline
 Memory (MB) & [100-500] \\ 
 CPU (MIPS) & [100-900] \\
 Storage (GB) & [1-6] \\
 Max Throughput (Mbps) & [10-50] \\
 \hline
\end{tabular}
\end{center}
\caption{Microservices requirements.}
\label{tab:MicroservicesRequirements}
\end{table}

In the simulations, we assume a time-slotted system, in which each time-step corresponds to some requests that the application had to process. Moreover, the application structure and the cloud infrastructure are fixed and unaltered among all the experiments. We determine the initial placement following the \textit{initial placement} procedure described in Section~\ref{sec:placement} and then run the \textit{DSR} algorithm. Finally, we set always a difference of 0.5 seconds between the maximum allowed completion time and the maximum processing time.

\subsection{Comparison between the heuristic and the solver}

\begin{figure*}[t]
\centering
\begin{subfigure}{.40\textwidth}
  \centering
  \includegraphics[width=1\linewidth]{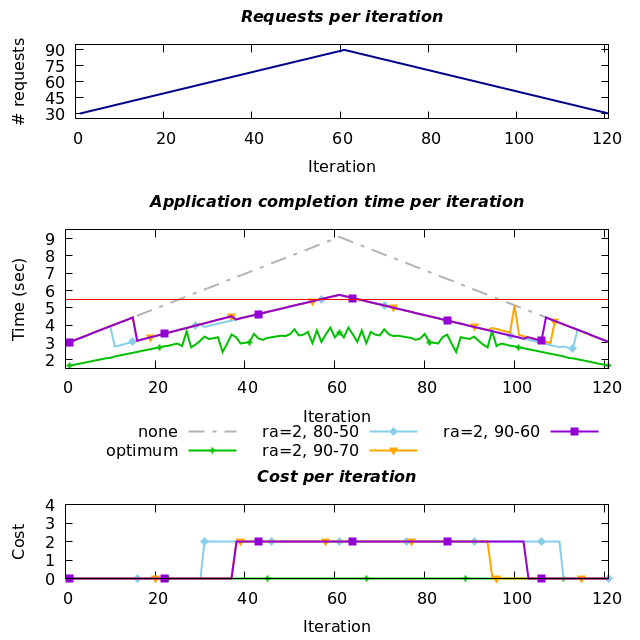}
  \caption{}
  \label{fig:2thresholds}
\end{subfigure}\hspace{5mm}
\begin{subfigure}{.40\textwidth}
  \centering
  \includegraphics[width=1\linewidth]{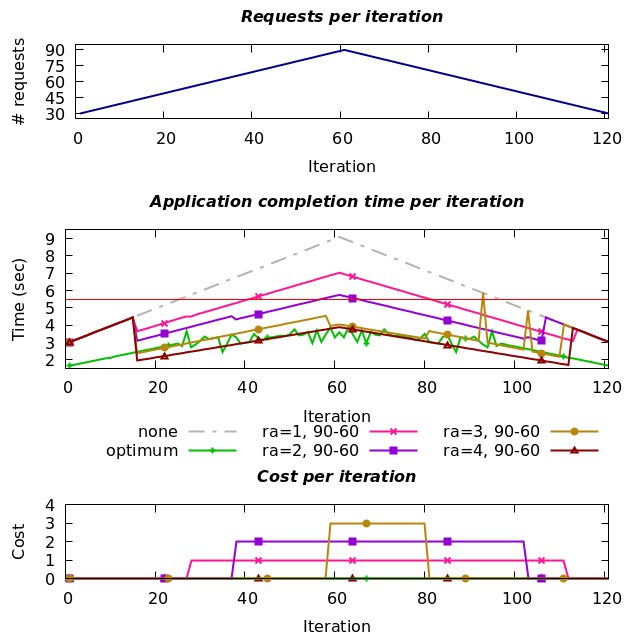}
  \caption{}
  \label{fig:9060dimensions}
\end{subfigure}
\caption{Application completion time and provider costs considering a monotonic increasing and then decreasing requests pattern; (a) fixed replicasArray's size different thresholds (b) fixed threshold different replicasArray's sizes.}
\label{fig:incDec}
\end{figure*}

In these experiments we mainly focus on the application completion time and the provider costs given by the heuristic algorithm and the mathematical model. The heuristic has two parameters that allow several configurations: the replicasArray size and the processing time thresholds. In the following simulations we indicate the former as `ra', and the latter as `y-z'. `y-z' means that, once the y\% of the allowed processing time has been exceeded, the \textit{DSR} algorithm tries to activate a new replicasArray; while, once the processing time falls below the z\%, it tries to deactivate a replicasArray as described in Algo.~\ref{alg:isWorkloadShiftRequired}. Finally, the maximum allowed completion time for the application is represented by the horizontal red line in the plots.  

Figure \ref{fig:incDec} considers a monotonic increase and then a monotonic decrease of the requests sent to the application. In Figure \ref{fig:2thresholds} we set the size of the replicasArray to 2, and we test different thresholds, 90-60, 90-70 and 80-50. In Figure \ref{fig:9060dimensions} we choose a threshold, 90-60, and we vary the replicasArray size between 1 and 4. Moreover, in both the plots, we represent the solution obtained without any redirection of the traffic to replicasArrays and only the one at the edge active, called `none'. In Figure \ref{fig:2thresholds} we can notice that all the heuristic solutions perform similarly in the completion time required to process the requests. The differences are in the activation and deactivation times; due to different thresholds, the redirection of the workload happens at different iterations. Peaks on the lines of the heuristics solutions, except the one at iteration 61, represent the effects of the activation/deactivation of a replicasArray. The solution that takes no decision about redirecting traffic after the initial placement can be seen that performs poorly in terms of completion time; most of the time, it exceeds the maximum allowed. When the \textit{DSR} algorithm has no residual replicasArrays activatable, time violations may happen, as it can be observed around iteration 60. 
In Figure\ref{fig:9060dimensions} we can see the effect of changing the replicasArray size. The differences in terms of the completion among the heuristics are now more evident; the performance improves with the increasing of the replicasArray dimension.
Finally, in both the plots of Figure \ref{fig:incDec}, we can also inspect the costs associated with each solution. The cost per iteration is given by the number of microservices that are receiving traffic, at that moment, in a public cloud. In fact, the cost of each heuristic solution depends on the replicasArray's size and the moment of activation and deactivation of the public region. The optimal solution never uses the public cloud to process the requests and thus the cost is always 0. In Figure \ref{fig:9060dimensions} we can observe the impact of varying the replicasArray size. Having more  active microservices in the replicasArray means adding more computational power to the application. However, this counteracts the cost: if some traffic is sent to a public region the cost will be greater due to a bigger number of microservices hosted there. It is important to find a good trade-off between performance and costs. A bigger replicasArray leads to lower completion time but at the cost of more resources required and greater renting costs. 

\begin{figure*}[t!]
\centering
\begin{subfigure}{.40\textwidth}
  \centering
  \includegraphics[width=1\linewidth]{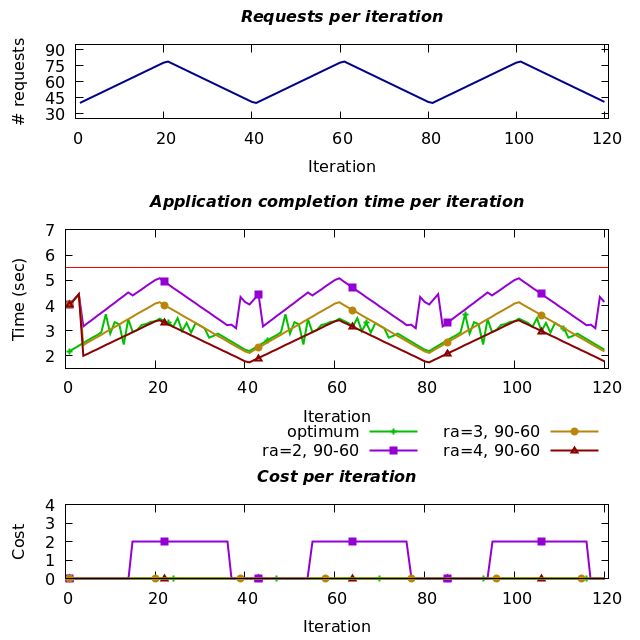}
  \caption{}
  \label{fig:periodic}
\end{subfigure}\hspace{5mm}
\begin{subfigure}{.40\textwidth}
  \centering
  \includegraphics[width=1\linewidth]{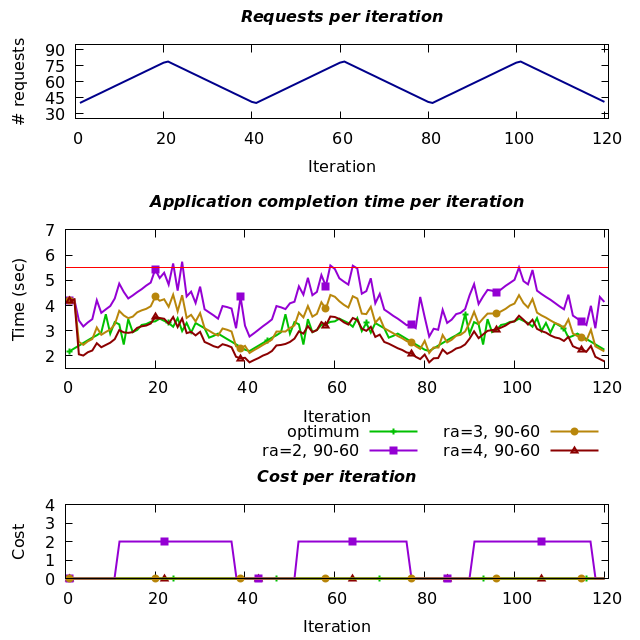}
  \caption{}
  \label{fig:periodic_noise}
\end{subfigure}
\caption{Application completion time and provider costs considering a periodic requests pattern; (a) without and (b) with external factors.}
\label{fig:periodic_comparison}
\end{figure*}

The experiments depicted in Figure \ref{fig:periodic_comparison} are conducted to test the proposed solution under different requests pattern and evaluate how external factors, such as the coexistence and concurrent execution of processes not related to the application or other exogenous factors due to the utilization of the cpu, influence the behaviour of the algorithm. In particular, that kind of noise is simulated by choosing an arbitrary region among the ones available and consuming a random amount, within the range [0-250] MIPS, of its CPU resources.

Figure \ref{fig:periodic_comparison} considers a periodic requests pattern sent to the application. Figure \ref{fig:periodic} shows how the solutions behave under the considered pattern in optimal circumstances. The experiment represented in Figure~\ref{fig:periodic_noise}, on the other hand, shows how the solutions operate, in the same scenario, but with external factors. Analyzing the plots, it can be noticed that the heuristic solutions having replicasArray size to 3 and 4 do not suffer much the external factors and perform similarly to the mathematical formulation in terms of completion time. Moreover, as the latter, both the heuristics never involve the public region. Instead, the heuristic solution with replicasArray dimension set to 2, under high load, involves the public resources. Moreover, the random noise causes some time violations and high peaks in the completion time curve; this is because the small replicasArray dimension and the consequently limited power computation leads to a higher sensitivity to external factors.

\subsection{Comparison with the State of the Art}

\begin{figure*}[t!]
\centering
\begin{subfigure}{.33\textwidth}
  \centering
  \includegraphics[width=\textwidth]{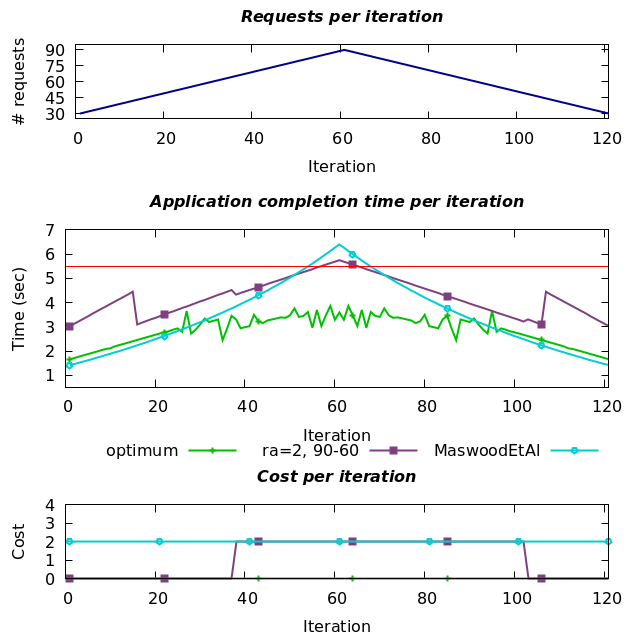}
  \caption{}
  \label{fig:MEA_half_incdec2}
\end{subfigure}%
\begin{subfigure}{.33\textwidth}
  \centering
  \includegraphics[width=\textwidth]{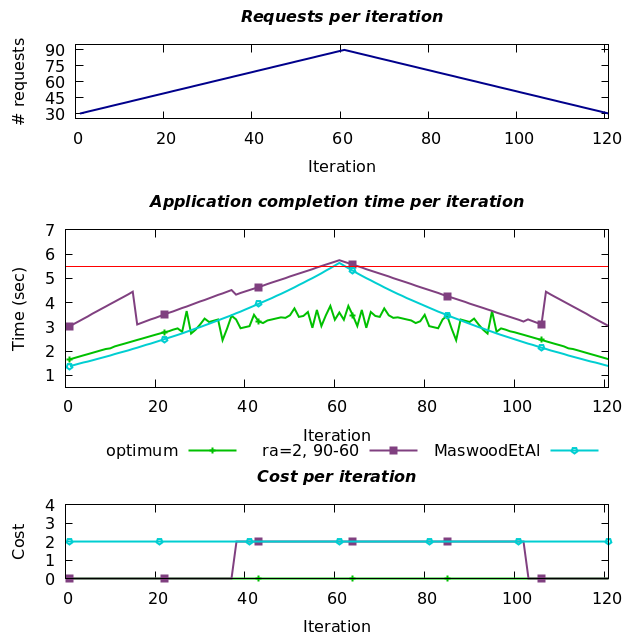}
  \caption{}
  \label{fig:MEA_lb_incdec2}
\end{subfigure}%
\begin{subfigure}{.33\textwidth}
  \centering
  \includegraphics[width=\textwidth]{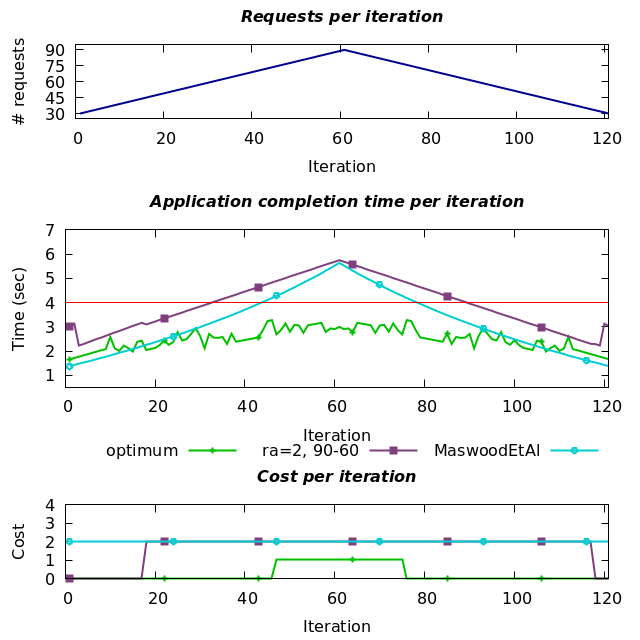}
  \caption{}
  \label{fig:MEA_4_incdec2}
\end{subfigure}
\caption{Application completion time and provider costs considering a monotonic increasing and then decreasing requests pattern; (a) prioritizing cost minimization and load balancing (b) prioritizing load balancing only (c) prioritizing load balancing only with a different maximum completion time.}
\label{fig:MEA_half_lb}
\end{figure*}

In these experiments, the comparison is done assuming that the regions hosting the replicasArrays are the servers in the three-layer fog-cloud computing environment. We do not take into account the goal out of our scope, i.e. minimize the maximum link utilization of network links, setting its weight, $\beta$, always to 0. Moreover, we do not consider the single servers within regions and we assume a single path between the regions. The total amount of resource demand and the total amount of bandwidth demand are computed depending on the number of requests intended for the replicasArrays. The former is calculated as the number of requests times $\omega_m$ for each $m$ in the replicasArray. The latter is computed as the number of requests times the average input requests size, $\Delta_m$. Finally, it must be noted that the assumption of single servers and single paths allows us lower weight parameters tuning. 

In the experiment depicted in Figure \ref{fig:MEA_half_lb}, we compare the proposed algorithms and the one presented in \cite{MaswoodEA_BandwidthLoadbalancing}. In Figure \ref{fig:MEA_half_incdec2} we set the MEA objective function to minimize the bandwidth costs while introducing load balancing. On the other hand, in Figure \ref{fig:MEA_lb_incdec2} and \ref{fig:MEA_4_incdec2} we set the MEA weight parameters to prioritize load balancing only. To obtain the former workload allocation strategy we set the weights to: $\alpha = 0$, $\beta = 0$ and $\gamma = 1$ while to obtain the latter to $\alpha = -1$, $\beta = 0$ and $\gamma = 1$. This is because, as stated in \cite{MaswoodEA_BandwidthLoadbalancing}, the maximum server resource utilization decreases with the increase in $\gamma$ and the demand starts to be distributed as we continue to decrease $\alpha$ compared to the other weight factors. Furthermore, in these experiments, we aim to show how the presented algorithms behave with a different maximum completion time threshold.

Figure \ref{fig:MEA_half_lb} depicts the behaviour of the solutions considering an increasing and then a decreasing requests pattern. The introduction of load balancing allows MEA, as depicted in Figure \ref{fig:MEA_half_incdec2}, to perform, in terms of completion time, mostly better than the heuristic. However, under high workload conditions, MEA is the solution that behaves worse in the time required to process the submitted requests and the load balancing aspect influences the provider costs. 

When the priority of MEA is set to load balancing only, as in Figure \ref{fig:MEA_lb_incdec2} and \ref{fig:MEA_4_incdec2}, it is always capable of lower application completion time with respect to the heuristic. This is achieved by MEA, if compared to Figure \ref{fig:MEA_half_incdec2}, performing a better load balancing and by an increasing involvement of the public region. However, by changing the maximum completion time threshold, although both MEA and the heuristic violate the time constraint, the \textit{DSR} algorithm reacts and utilizes the replicasArrays differently, while MEA, that has not any QoS, performs the same. The self-adaptation is visible looking at the peaks in the line of the completion time and from the cost plot; the heuristic, trying not to pass the threshold, uses, and pays, for many more iterations the public resources. Furthermore, it can be seen that also the optimal solutions needs to use the public region to avoid the violation of the new time constraint. 

From these simulations, it can be understood that MEA does not react to a new time threshold and does not violate the constraint only if this is set to a high enough value. This is because MEA has no QoS. On the other hand, the heuristic and the optimal solution can adapt themselves and, subject to higher costs, try not to violate a more stringent constraint.

\subsection{Real costs simulations}
\label{subsec:real_costs_sim}

In the following experiment, we aim to evidence the possible real costs associated with the adoption of the presented algorithms. We compute them leveraging some previous scenarios and Amazon AWS Lambda \cite{amazon_AWS_lambda}. We simulate the usage of Amazon AWS Lambda since it allows a pay-per-use policy. AWS users are charged based on the number of requests for their code and depending on the execution time. The price depends on the amount of memory allocated to the code. These features match with our assumptions; a provider pays depending on the size and the usage of a replicasArray, hosted in a pubic region, that is ready to receive requests. 

All the previous simulations span 120 iterations; assuming that each time slot lasts for one minute a simulation covers 2 hours. Thus, considering that a scenario repeats itself once completed, by multiplying the output of an instance times the pairs of hours in a month, 360, we can obtain the monthly costs. The reference prices for resources in the Europe(Milan) region is 0.0000195172\$ for every GB-second. Assuming that a replicasArray of size two requires 1024 MB of memory and its execution lasts 1 second, the total costs can be computed as:
\begin{align}
    tot\_req &= req\_to\_public\_in\_one\_simulation * 360 \nonumber \\
    tot\_sec &= tot\_req * 1 sec \nonumber \\
    tot\_GB/s &= tot\_sec * (1024 / 1024)  \nonumber \\
    tot\_costs &= tot\_GB/s * 0.0000195172 \nonumber
\end{align}
\begin{figure*}[!t]
\centering
\begin{subfigure}{.35\textwidth}
  \centering
  \includegraphics[width=1\linewidth]{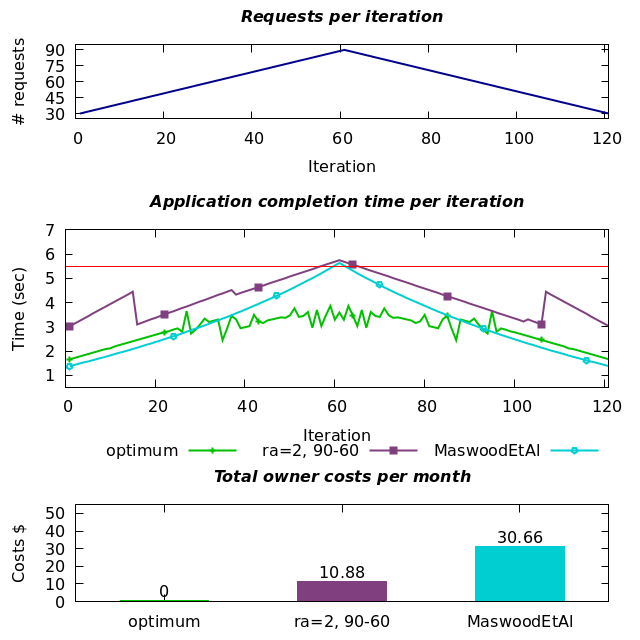}
  \caption{}
  \label{fig:MEA_amazon_5incdec2}
\end{subfigure}\hspace{5mm}
\begin{subfigure}{.35\textwidth}
  \centering
  \includegraphics[width=1\linewidth]{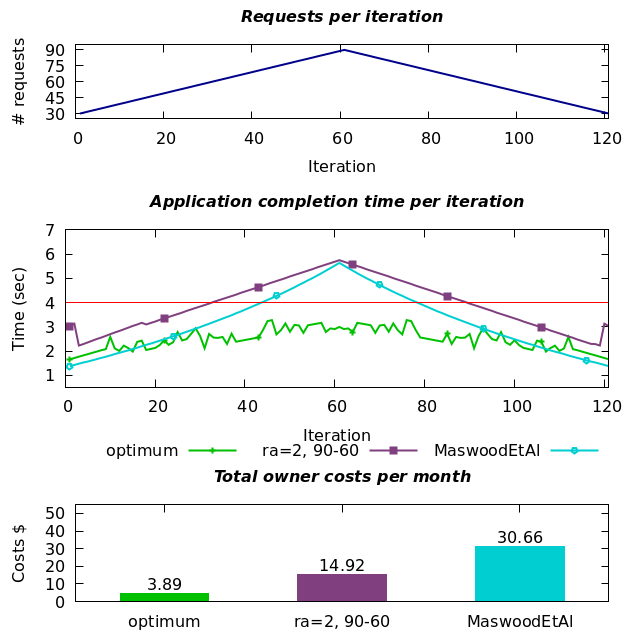}
  \caption{}
  \label{fig:MEA_amazon_4incdec2}
\end{subfigure}
\caption{Application completion time and provider costs considering a monotonic increasing and then decreasing requests pattern prioritizing load balancing; (a) maximum completion time 5.5 sec (b) maximum completion time 4 sec.}
\label{fig:MEA_amazon_indec}
\end{figure*}

Figure \ref{fig:MEA_amazon_indec} depicts the performance of the solutions considering an increasing and then a decreasing requests pattern. Figure \ref{fig:MEA_amazon_5incdec2} considers a maximum completion time of 5.5 seconds and refers to Figure \ref{fig:MEA_lb_incdec2}, while \ref{fig:MEA_amazon_4incdec2} of 4 seconds and is related to Figure \ref{fig:MEA_4_incdec2}. Analyzing the two plots, it can be noticed that MEA does not change its behaviour and thus reaches the same costs. Conversely, the mathematical formulation and the heuristic adapt themselves and, due to more stringent constraints, pay more. However the latter, although it performs worse in terms of completion time, allows the provider to save more than 50\% of the costs, in both the simulations, compared to MEA.

\section{Real Implementation}\label{sec:impl}
In this section, we describe the implementation of the proposed heuristic and the results obtained in a real cloud environment. As a use case scenario, we introduce the PaperMiner application. 

\subsection{PaperMiner application}
\label{sec:PaperMiner_app}
\begin{figure}[t]
\includegraphics[width=0.45\textwidth]{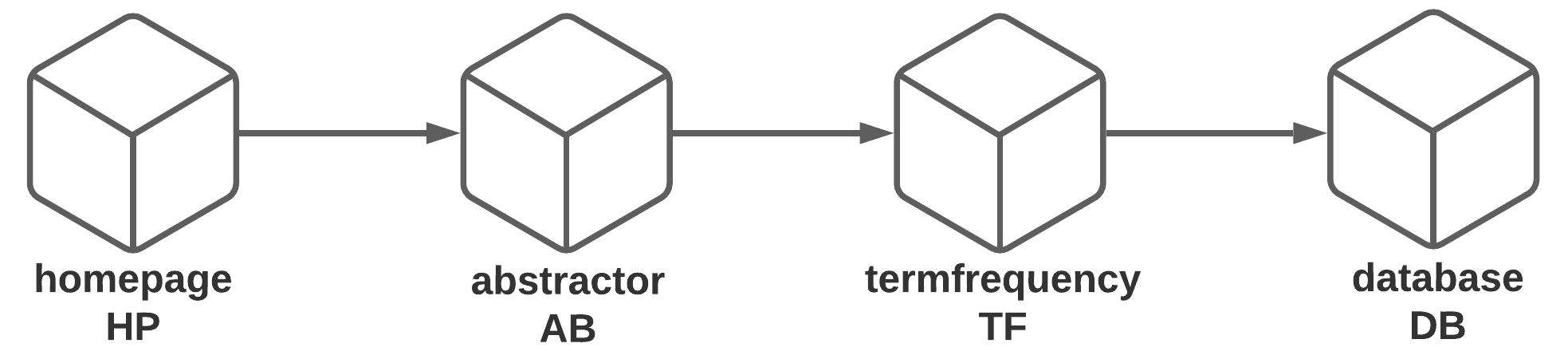}
\centering
\caption{PaperMiner architecture diagram.}
\label{fig:paperminer}
\end{figure}

The application presents a homepage that allows anonymous users to upload the PDFs of scientific papers that they want to “mine”. Once the files are uploaded, the application automatically extracts, for each of them, its abstract and its term-frequency statistic. The results are then stored in a persistent database and a confirmation message is returned to users. PaperMiner is built following the microservice architecture paradigm and is composed of four microservices, as depicted in Figure \ref{fig:paperminer}. These, except the database, are developed as Python programs following the asynchronous request-reply pattern. We leverage the AIOHTTP \cite{AIOHTTP} library that allows the microservices to be non-blocking, i.e., a request does not block the client until the operation completes. For the storage, we utilize a document-based database such as mongoDB \cite{mongodb}.

\subsection{System architecture}
\label{sec:cloud_infrastructure}
The hybrid cloud scenario has been created as a Kubernetes cluster in an OpenStack \cite{openstack} ecosystem. The nodes configured, Virtual Machines that run Ubuntu 16.04.3 LTS, are four and they are subdivided into one Master and three Workers nodes. We deploy the same infrastructure of the simulation phase, two private regions and a public one, and let each worker node represent a different region of the cloud. In details, the nodes have the features described in Table \ref{tab:nodeFlavor}.

\begin{table}[t]
\begin{center}
\begin{tabular}{ |c|c|c|c|c| } 
 \hline
 Node & Region & vCPU & RAM (GB) & HDD (GB)\\ \hline
 Master & / & 2 & 2 & 40 \\ 
 Worker-edge & Edge & 2 & 4 & 40 \\
 Worker-central & Central & 4 & 8 & 40 \\
 Worker-public & Public & 4 & 8 & 40 \\
 \hline
\end{tabular}
\end{center}
\caption{Cluster's nodes flavor.}
\label{tab:nodeFlavor}
\end{table}
Nodes are configured in order to capture that edge resources have constrained computation capabilities while the central and the public region have greater computational capabilities.

\subsection{Placement}
Following the \textit{initial placement} algorithm defined previously, we define the subset of microservices that compose the replicasArrays and the initial placement of the PaperMiner application. There is a one-to-one mapping between regions and worker-nodes, as reported in Table \ref{tab:nodeFlavor}. The replicasArray cosists of two microservices, abstractor and termfrequency, and the initial assumption is that the incoming traffic is processed by the edge replicasArray only. 

\begin{figure*}[t]
\centering
\begin{subfigure}{.40\textwidth}
  \centering
  \includegraphics[width=1\linewidth]{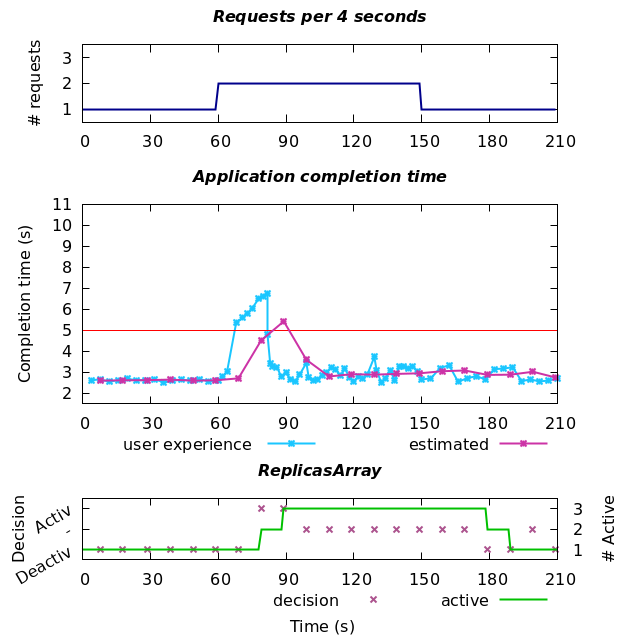}
  \caption{}
  \label{fig:realCloud_const_UpDown}
\end{subfigure}\hspace{5mm}
\begin{subfigure}{.40\textwidth}
  \centering
  \includegraphics[width=1\linewidth]{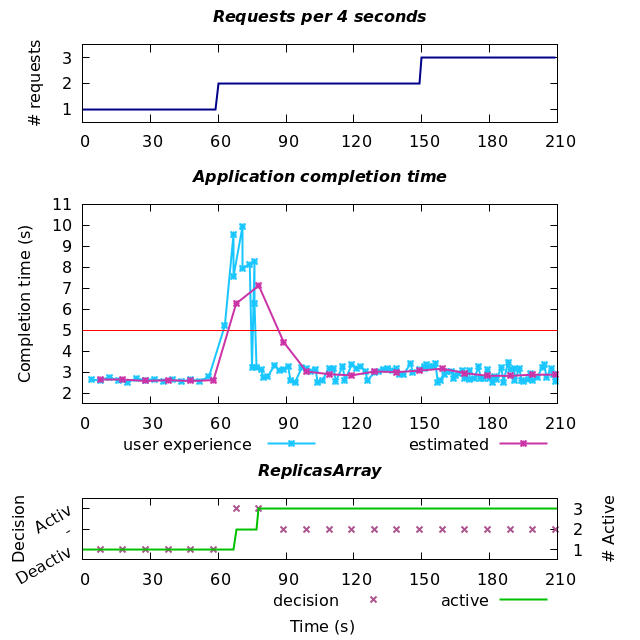}
  \caption{}
  \label{fig:realCloud_const_Up}
\end{subfigure}
\caption{Application completion times and replicasArray active considering three intervals of a constant requests pattern.}
\label{fig:realCloud}
\end{figure*} 

\subsection{Network}
After the creation of a virtual network topology leveraging OpenStack, since Kubernetes assumes a unique IP address for each pod within the cluster, we deploy flannel \cite{flannel} as container network interface. Flannel is a virtual network that gives a subnet to each host. Finally, on top of flannel, we build a service mesh installing Istio \cite{Istio}. Istio injects proxies alongside each service of the application in order to handle inter-service communications. All the microservices are packaged in a pod with a sidecar proxy that intercepts incoming and outgoing calls for the services.

This allows us to implement the custom network functionalities desired. By leveraging `virtual services', a key building block of Istio’s traffic routing functionality, we are able to configure how requests are routed within the service mesh. Firstly, we create one `virtual service' to control the quantity of traffic that each replicasArray receives. Then, we define as many `virtual services' as the number of replicasArray to regulate the communication among microservices of the same replicasArray. Internal replicasArray requests can not be outsourced.

\subsection{Monitoring}
We leverage Prometheus \cite{prometheus} to store and retrieve metrics about the state of the cluster. The Prometheus server is configured to run in a pod and to be hosted in the private central region. The custom configuration we deploy has a scrape interval set to five seconds and it collects metrics from two sets of targets. The first ones are the service mesh sidecar proxies; Istio, by default, defines and generates a set of standard metrics. These provide useful information about the service mesh, like the request duration and the request number. The second set of targets are the worker nodes. These are instrumented, through the `node\_exporter' software, to expose hardware and OS metrics, like the CPU statistics.

\subsection{Evaluation}
To test the performance of the proposed heuristic we simulate, using the load testing tool k6 \cite{k6}, anonymous users sending a request containing a scientific paper in PDF format.

The maximum completion time is set to five seconds and the upper and lower heuristic thresholds are 90 and 60, respectively. In Figure \ref{fig:realCloud}, a point on the `user experience' line represents the completion time of a request measured and reported, at its arrival time, by the load testing tool. Instead, a point on the `estimated' line is the result of a Prometheus query. The `estimated' points are meant to capture the trend of the application completion time and they are computed as the average of the values within the evaluation interval. They are used by the \textit{DSR} algorithm to decide about the redirection of the workload. Finally, a point in the ReplicasArray sub-chart represents the decision taken by the \textit{DSR} algorithm about the redirection of the traffic while the line keeps track of the number of replicasArray active.

In the experiment depicted in Figure \ref{fig:realCloud_const_UpDown}, we simulate a requests pattern made of three intervals of constant load. As it can be seen, in the first minute of the simulation, a request every four seconds is sent. Then, for the following ninety seconds, the requests sent in the time-unit, four seconds, duplicates. Finally, in the subsequent minute, the requests sent have the same frequency of the first interval, one every four seconds. The first period of the experiment is defined to simulate a scenario where the number of requests sent can be processed only by edge resources and without any saturation of that region. Indeed, in optimal conditions, the average completion time of requests containing the PDF of the scientific paper is about 2.8 seconds. These remarks are confirmed by the application completion time plot. Analyzing the `user experience' line, it can be noticed that all the requests of the first interval are processed in approximately the same amount of time. The second period of the experiment, from the 60th to the 150th second, simulates a high load that can not be processed only by edge resources. This can be observed by analyzing the completion time. Users, the `user experience' line, start from the 60th second to experience an increasing completion time to their requests. This behaviour continues until the \textit{DSR} algorithm, at the 79th second, detects a threshold violation and decides to activate a region. This is visible analyzing both the application completion time and the replicasArray plot. In fact, it can be seen in the latter that the decision is `activate' and the number of replicasArray transitions from one to two. The effect of the redirection of the workload is noticeable by users some seconds after when the experienced completion times return to acceptable values. The `estimated' line takes some seconds to capture the trend of the user experience due to the scrape and evaluation interval of Prometheus, and the ten seconds pause between the iterations of the infinite loop. At the 89th second, all the available replicasArray are active and receiving traffic. The redirection of the traffic lets the algorithm detect, from the 99th to the 169th second, that the requests complete, on average, within the threshold values defined and the decision is to do not change the current network configuration. From the 150th second, the requests submitted return to a value that potentially allows only the edge region to process them. In fact, the heuristic reaches this configuration deactivating, at the 179th second, the public region and then, at the 189th second, the central one. 

In the experiment represented in Figure \ref{fig:realCloud_const_Up}, we simulate a requests pattern made of three intervals of an increasing constant load. As it can be noticed, the first two intervals are the same as the previous simulation but, in the last minute of this run, the requests sent in the time-unit increase to three. As can be seen in the plot, in the first period of the experiment the requests are processed in a satisfactory manner using edge resources only. Since there is a single replicasArray active, the `deactivate' decisions have no effect and the network configuration does not change. Then, at the 60th second, the requests sent duplicate and the users start to experience an increasing completion time to their requests. This is detected by the \textit{DSR} algorithm that, at the 68th and 78th seconds, activates one region at a time. The decisions taken by the algorithm after the transition from one to two requests per time-unit look very similar to the decisions taken in the experiment depicted in Figure \ref{fig:realCloud_const_UpDown} at the corresponding moments; this confirms the good performance of the proposed solution. From the 78th second, all the replicasArray are active and receiving traffic; this lets the heuristic detect, from the 88th second until the end of the simulation, that requests complete, on average, within the threshold values defined and no changes are required. During the third interval, the \textit{DSR} algorithm does not decide to deactivate, as in the previous experiment, since the required conditions are not matched because of the increase in the frequency requests. This confirms the validity of the solution proposed that is able to elastically handle and respond to the different workload situations. Furthermore, it can be noticed, from the 150th second, that the heuristic efficiently load balances the replicasArrays available and provides stability even under high workload conditions; the variance in the completion time experienced by users is very limited.

\section{Conclusion}
In this paper, a cost-effective workload allocation strategy for fog cloud-native edge computing services is studied. Firstly, we modeled the system considered and then we proposed an integer programming to formulate the problem under investigation. Given its high computational complexity, we presented a heuristic algorithm that addresses the problem of workload allocation by redirecting the data flow from one cloud region to another. The aim was to reduce, for a cloud service provider, the costs of the public cloud and to meet a set of predefined constraints. The algorithm, leveraging the concept of replicasArray, is able to efficiently distribute the traffic within the cluster. In fact, having a single point of splitting and by load-balancing the microservices depending on the residual host capacity has been shown, by this work, to be effective.

We investigated several scenarios and the parameters that influence the heuristic. Moreover, we compared the proposed algorithm to an optimum and a literature solution. The optimum provides a lower bound to the provider's costs while the literature solution was used to understand the performance achievable when the priority is load-balancing. The results obtained evidence that the heuristic, in most of the scenarios where real costs were studied, can save more than 50\% of the load-balancing focused solution costs. However, this comes with a slight decrease in the application completion time performances; the heuristic approach and the cost-reduction focus impact the results. Nevertheless, the proposed algorithm presents good performance in almost all the situations studied and, subject to a large enough replicasArrays dimension, can respect the constraints and respond to requests pattern variation. With the appropriate quantity of computation resources instantiated, the heuristic represents a cost-effective solution able to fit and respond to different environments. Finally, we have proven that, leveraging some of the most recent technologies in the field of cloud computing, the heuristic can be deployed in a real cluster. The solution efficiently detects the load trend and takes the proper decision to respect the time constraint; public resources are utilized only for the time required to guarantee a proper user experience. Thus, we concluded that the proposed heuristic represents a viable cost-effective workload allocation strategy for cloud-native edge services. 

\subsection{Future Work} 
The replicasArray size affects many aspects, such as resources occupation and performance, of the proposed workload allocation strategy. Currently, defining the appropriate replicasArray dimension is a task left to cloud service providers. Moreover, this size can not be changed at run-time. In future works, we plan to extend the proposed solution with a procedure able to identify the proper dimension and to dynamically adjust the size, and deployment, of replicasArrays depending on demands.

 \section*{Acknowledgments}

This work has received funding from the European Union’s Horizon 2020 Research and Innovation Programme under grant agreement no. 815141 (DECENTER: Decentralised technologies for orchestrated Cloud-to-Edge intelligence).

 \bibliographystyle{elsarticle-num} 
 \bibliography{cas-refs}





\end{document}